# A Structure-based Memory Maintenance Model for Neural Tracking of Linguistic Structures

Nai Ding

Zhejiang University

**Abstract:** It is recently demonstrated that cortical activity can track the time courses of phrases and sentences during speech listening. Here, we propose a plausible neural processing framework to explain this phenomenon. It is argued that the brain maintains the neural representation of a linguistic unit, i.e., a word or a phrase, in a processing buffer until the unit is integrated into a higher-level structure. After being integrated, the unit is removed from the buffer and becomes activated long-term memory. In this model, the duration each unit is maintained in the processing buffer depends on the linguistic structure of the speech input. It is shown that the number of items retained in the processing buffer follows the time courses of phrases and sentences, in line with neurophysiological data, whether the syntactic structure of a sentence is mentally parsed using a bottom-up or top-down predictive model. This model generates a range of testable predictions about the link between linguistic structures, their dynamic psychological representations and their neural underpinnings.

**Introduction**

During speech processing, how the brain integrates information across words to derive the meaning of a sentence is a fundamental question in cognitive neuroscience (Chomsky, 1965; Townsend and Bever, 2001). This question has been addressed using a variety of approaches, including theoretical modeling (Frazier and Fodor, 1978; Berwick and Weinberg, 1986; Prince and Smolensky, 1997; Phillips, 2003; Lewis et al., 2006; Frank et al., 2012; Martin and Doumas, 2017) behavioral experiments (Fodor and Bever, 1965; Garrett et al., 1966; Marslen-Wilson et al., 1978; Marslen-Wilson and Tyler, 1980; Frazier and Rayner, 1982; Trueswell et al., 1994; Ferreira et al., 2002; Grodner and Gibson, 2005), and neural experiments (Kutas and Federmeier, 2000; Friederici, 2002; Pallier et al., 2011; Brennan et al., 2012; Bemis and Pylkkänen, 2013; Brennan et al., 2016). Recently, a number of neurophysiology and neuroimaging studies suggest that ongoing neural activity can track the constituent structure of a sentence (Bastiaansen et al., 2010; Pallier et al., 2011; Peña and Melloni, 2012; Ding et al., 2016;

Meyer et al., 2016; Nelson et al., 2017). These studies have reported either sustained neural activity or consistently changing neural activity over the time course of a phrase or sentence. Such constituent-tracking neural activity provides neural evidence that the brain hierarchically groups words into phrases and sentences (Ding et al., 2016). Mechanistically, however, it remains controversial what are computations generating constituent-tracking activity.

At the linguistic level, it has been suggested that constituent-tracking neural activity reflects information integration within constituents (Pallier et al., 2011), syntactic or semantic unification between words (Bastiaansen and Hagoort, 2015), or syntactic parsing (Nelson et al., 2017). At the level of neural implementations, it is been proposed that sustained activity during a sentence plays a role in coordinating neural ensembles that encode different components of a sentence (Peña and Melloni, 2012; Meyer et al., 2016). Furthermore, it is recently shown that a model initially developed for relational reasoning produces constituent-tracking responses when applied to speech processing (Martin and Doumas, 2017).

Here we propose a model that relates the constituent-tracking neurophysiological responses with memory-related operations. It is argued that the neural tracking of syntactic structures may be related to the number of linguistic units stored in working memory. The relatoinship between language processing and working memory has been extensively discussed, mostly focusing on complex sentences with long-distance dependencies that challenge the working memory capacity (Daneman and Carpenter, 1980; Just and Carpenter, 1992; Gibson, 1998; Caplan and Waters, 1999; McElree et al., 2003; Lewis et al., 2006; Christiansen and Chater, 2016). The current model, however, focuses on neural processing of local dependency. Since working memory is a broad concept and here we only discuss the part of working memory related to linguistic structure building, in the following we will refer to this component of working memory as a processing buffer. The general hypothesis is that only a limited number of linguistic units can be kept in the processing buffer and therefore the brain constantly updates the processing buffer and keeps only linguistic units that may be immediately used for subsequent linguistic structure building (Fig. 1, criterion A). In the following, we futher develop the hypothesis into a structure-based memory maintenance model (Fig. 1, criterion B).

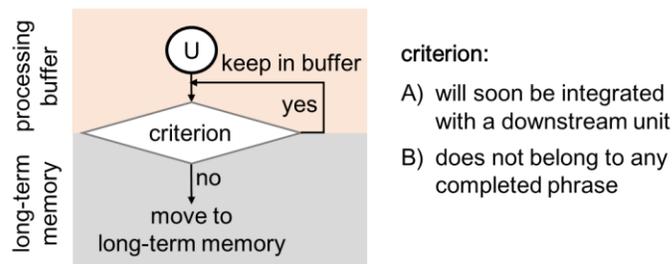

Figure 1. Schematic illustration of the structure-based memory maintenance model. The processing buffer has limited capacity and holds linguistic units that can be directly used for linguistic structure building. The long-term memory, in contrast, is not limited in capacity but costs time to access. Units held in the processing buffer are constantly updated and only units that are most likely to be immediately used for subsequent linguistic structure building will be kept in the buffer (criterion A). Specifically, it is assumed that only linguistic units that have not been integrated into a superordinate structure is likely to be used in subsequent linguistic structure building (criterion B).

**Structure-based Memory Maintenance Model**

Consider a short phrase, e.g., "fun games", that is presented auditorily. The listener first hear "fun" and then "games". To integrate these two words, the brain must have access to the neural representations of both words. This means that neural integration of these words cannot start until the second word is heard and when integration occurs the neural representation of the first word must be active. Here, in the structure-based memory maintenance model, we hypothesize that the neural representation of the first word is actively retained in a processing buffer until it is integrated with the second word. This is in contrast to models in which each previous word is stored in active long-term memory and reactivated by retrieval when a subsequent word needs to be integrated with previous words (Lewis et al., 2006).

After two words, e.g., "fun" and "games", combine into a phrase, the phrase as a whole often becomes the unit to integrate with other elements in a sentence. For example, in the sentence "fun games waste time", the predicate "waste time" integrates with "fun games", rather than "fun" or "games" per se. Therefore, the brain only needs a representation of "fun games" but not the representations of its component words when processing the second part of the sentence. The processing buffer is assumed to have high processing speed but limited capacity. Therefore items that are not potentially useful for subsequent linguistic structure building should be removed from the processing buffer. Therefore, neural representations of subordinate linguistic units, such as "fun" and "games", should be moved from the processing buffer to the activated long-term memory, which can hold a large number of items but is only accessible via cue-based search (McElree et al., 2003; Lewis et al., 2006).

To summarize, the structure-based memory maintenance model assumes that the neural representation of a linguistic unit, i.e., a word or a phrase, is maintained in the processing buffer until it is integrated into a higher-level linguistic structure. If the linguistic structure is represented by a tree structure, the working hypothesis can be reformulated as the following: In a syntactic tree, a node remains active, i.e., stored in the processing buffer, until it has an active parent node.

Based on this working hypothesis, the activation time courses of different linguistic units are shown in Fig. 2ab, for a simple sentence "fun games waste time". This model represents only one possible way to parse the sentence and is called the bottom-up model in this article. In Fig. 2a, a linguistic unit, word or phrase, is not integrated into a higher-level structure until all components of the higher-level structure are represented in the brain. Therefore, this model does not process sentences in an incremental way, in the sense that a word, e.g., "waste", may be buffered and not incrementally attached to the currently available structures. This issue may become more apparent when processing sentences with a right-branching structure (Fig. 3a). An alternative way is to create transient structures such as [[fun games] waste] for each word and adjust the structure when subsequent words come (Phillips, 2003).

Language processing is known to be highly predictive but prediction is not considered in the bottom-up model in Fig. 2ab. An alternative processing model for the same sentence is shown in Fig. 2ef, in which the brain actively generates predictions for the possible structures of the unfolding sentence. For example, upon hearing "fun", the brain expects a noun to follow and combine with "fun" to form a noun phrase. The noun phrase will further combine with a verb phrase to form a sentence. Such predictions are generated based on the statistics summarized from previous language experience. For example,

"fun" could be an adjective or a noun but it is most likely an adjective and an adjective is most likely to be followed by a noun. Similarly, if a verb phrase is most likely constructed by a verb and an object, the brain will also assume a verb-object structure for the verb phrase.

The expectation for each item is updated over time until the item actually comes. For example, after the appearance of the word "fun", a verb can be expected in a later part of the sentence but it is difficult to pinpoint which specific verb may come. After the word "games", the search space for the subsequent verb is narrowed. The expected item opens a slot in the processing buffer and will be gradually filled with information. Here, we assume that the activation of an expected item increases when the expectation become more precise. In other words, if no particular word can be expected, the expected item will show little activation. In contrast, when a specific word is expected, the expected item will be strongly activated. For example, the object of the verb "waste" is highly predictable and therefore it will have relatively strong activation even before the word is heard. If the incoming word does not match with the brain's expectation, the brain will adjust its prediction, which is more apparent for the more complex example shown in Fig. 3b.

In the structure-based memory maintenance model, a linguistic unit is removed from the processing buffer once it is integrated into a higher-level structure. The underlying assumption is that the resulting higher-level structure as a whole will be integrated with other constituents in a sentence. This assumption, however, is not always true. For example, consider the sentence "I like red apples but not green ones". The last word in the sentence, i.e., "ones", refers to the word "apples" rather than the phrase "red apples". In this case, the brain has to retrieve the word "apples" and integrate it with "green". A solution to this problem is provided by the cue-based retrieval model (Lewis et al., 2006). The general idea is that after a word/phrase is removed from the processing buffer it is still stored in activated long-term memory and can be retrieved by cue-based search. For example, when processing "ones" in the sentence above, the brain will search for the noun it refers to. The search will return all possible candidates and the brain will choose the most feasible one from the candidates. This strategy is also useful when the brain makes a mistake during an initial analysis of a sentence, e.g., when analyzing garden-path sentences (Townsend and Bever, 2001).

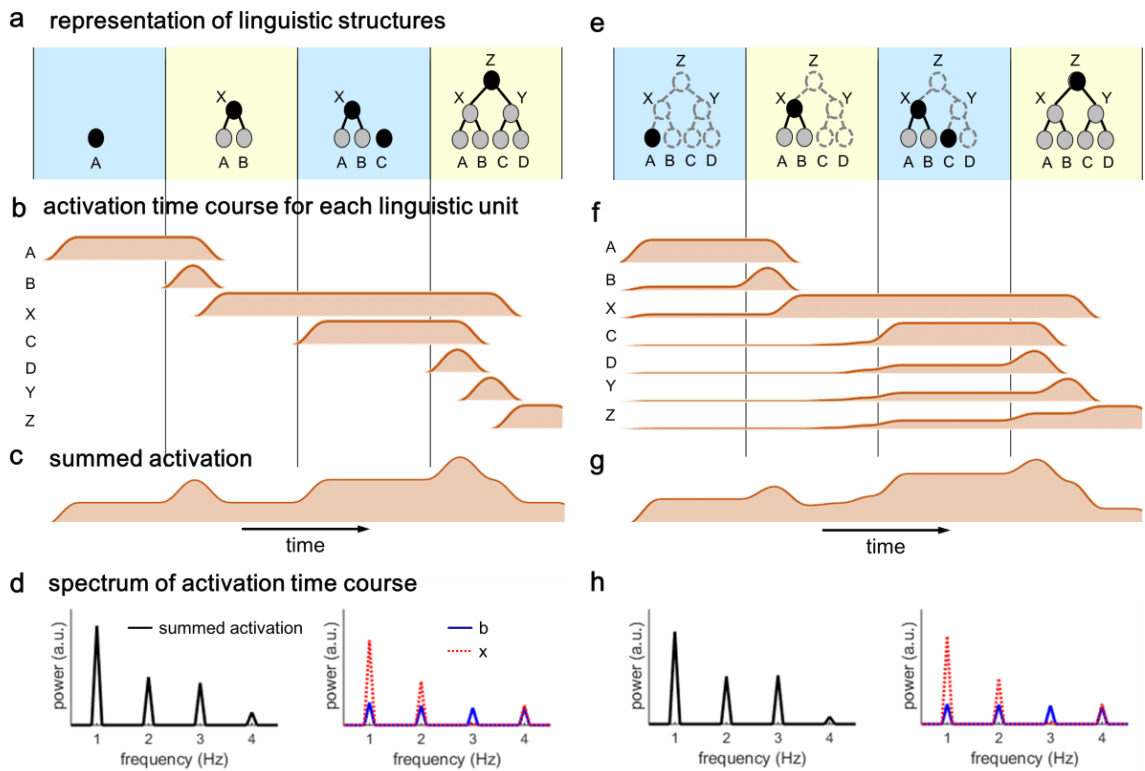

**Figure 2.** Evolution of the representations of different linguistic units in the processing buffer, for the sentence "fun games waste time". Panels a-d rely on a bottom-up parsing mechanism while panels e-h rely on a predicative top-down parsing mechanism. *Panels a & e:* The mental representation of the speech linguistic structure after hearing each word. The black nodes are maintained in the processing buffer while the gray nodes are stored in activated long-term memory. The hollow nodes with dashed lines indicate expected nodes, which are partially activated. The figures only show the final linguistic structure being derived after hearing each word, not showing the derivation process. Labels A-D represent the four words "fun", "games", "waste", "time", and labels X-Z represent phrases and sentences. *Panels b & f:* Time course of the neural representation of each linguistic unit in the processing buffer. *Panels c & g:* The sum of neural activation time courses in panels b & f. The summed time course can be viewed as the neural response recorded by a technique that cannot spatially resolve the neural representation of different linguistic units. *Panels d & h:* The power spectrum of the summed neural activation in panel b & f, and the power spectra of the activation of two representative node, i.e., node b & x. In this simulation, each word is set to 250 ms in duration and 10 sentences are sequentially played with no gap in between.

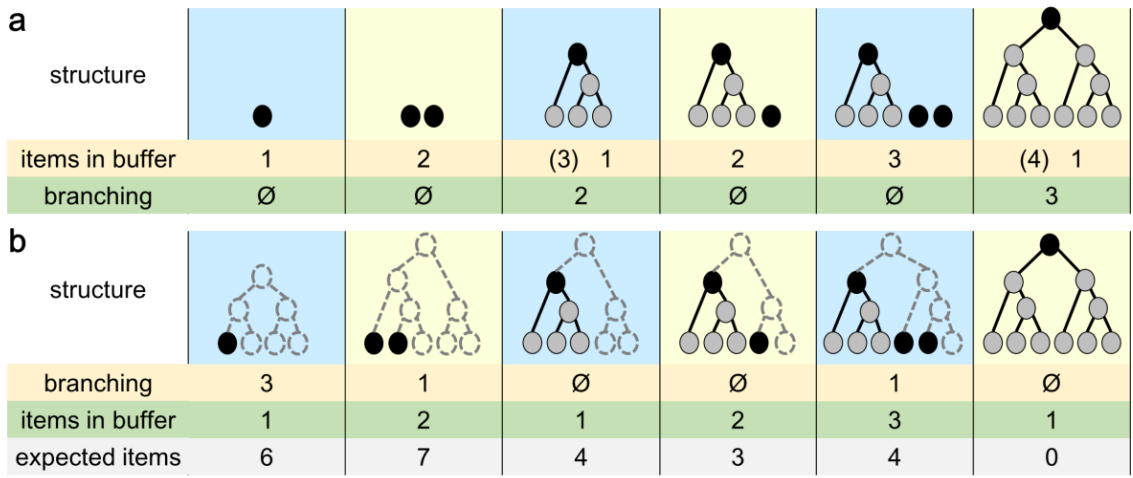

**Figure 3.** Evolution of the representation of a sentence "the little boy lost his keys" in the processing buffer, based on either a bottom-up parsing mechanism (a) or a predicative top-down parsing mechanism (b). The number of items retained in the processing buffer and the number of branching operations, i.e., attaching two child nodes to a node, are also shown.

**Simulation of previous MEG results**

In the following, we analyze if the structure-based memory-retrieval model can simulate MEG responses recorded during speech listening (Ding et al., 2016). Suppose that noninvasive neurophysiological data measured by MEG reflect the overall activation level of items in the processing buffer, i.e., the sum of activation of individual linguistic units (Fig. 2cg). In the simulation, the duration of each word is set to 250 ms and the stimulus is a sequence of 10 sentences with the same structure, i.e., [[adjective noun][verb noun]]. There is no acoustic gap between sentences. A frequency-domain analysis suggests that the overall activation level shows responses at 1 Hz and its harmonics (Fig. 2dh, left), consistent with previous MEG results (Ding et al., 2016). The simulation result is only one potential source of the MEG responses to sentences. For example, low-level auditory tracking will also generate a 4-Hz response, and the spectra of the activation time courses of individual nodes differ in their power at 1, 2, and 4 Hz (Fig. 2dh, right).

**Discussion**

The structure-based memory maintenance model provides a possible mechanism to manage linguistic items retained in the processing buffer for linguistic structure building. Here, the working definition of the processing buffer is a collection of neural representations that can be directly accessed during linguistic structure building. It can be interpreted as the workspace for linguistic structure building, which maintains a short list of items that the brain has quick access to, in contrast to a large pool of items in the long-term memory that can only be retrieved by a search. The processing buffer is a form of working memory. Nevertheless, since it is controversial whether the memory system used for language processing is the domain-specific or domain-general (Daneman and Carpenter, 1980; Caplan and Waters, 1999; Gordon et al., 2002; McElree et al., 2003), we do not directly link the processing buffer with a specific working memory theory.

**Assumptions underlying the structure-based memory maintenance model**

The primary assumption for the structure-based memory maintenance model is that a linguistic unit, word or phrase, must have an active neural representation when being integrated with other linguistic units. This assumption is intuitive but it does require the processing buffer to simultaneously handle more than one active items. More critically, it is assumed that, to process language efficiently, the neural representation of a linguistic unit should be kept active if the unit is expected to be integrated with subsequent linguistic units in the near future. The current model further hypothesizes that only linguistic units that have not been integrated into a higher-level linguistic structure are likely to be directly integrated with subsequent linguistic units. Therefore, only linguistic units that are parts of a incomplete dependency should be kept in the processing buffer. This assumption implies that items in the processing buffer do nonuniform decay time, i.e., some units being kept longer than others (Fig. 2). Indeed, recent working memory studies show that the memory for task-irrelevant features decays much faster than the memory for task-relevant features, even when both features are attended to during encoding (Chen and Wyble, 2015).

The current model puts forward a general framework but its details have to be specified by further empirical and theoretical work. For example, the current model does not specify the capacity of the processing buffer and how the brain remove less useful items when the buffer is short of storage space. It remains controversial how many items can be stored in working memory and how many items can be attended to simultaneously (Miller, 1956; Cowan, 2001; McElree, 2001; Ma et al., 2014). More critically, each item in the working memory is a chunk (e.g., a tree structure in the context of language processing) and multiple chunks may further construct a coherent scene (e.g., various parts of the same sentence that have not yet been integrated into a single tree structure). Therefore, before the organization of linguistic units in working memory is revealed, it is difficult to define how many linguistic units can be hold in working memory (Townsend and Bever, 2001).

Another question is what is maintained in the processing buffer. Here, we adopt the notion using by Lewis and colleagues (Lewis et al., 2006) and assume that the items in the processing buffers are represented as bundles of features. We do not specified, however, whether the processing buffer holds the actual spatial-temporal neural representation of a linguistic unit or a "pointer" to the spatial-temporal neural representation.

**Relationship with other dynamic language processing models**

The structure-based memory maintenance model is related to many other language processing models that describe how neural or psychological processes track linguistic structures over time. In the following, we compare the current model with previous models: In the information integration model proposed by Pallier and colleagues (Pallier et al., 2011), there is monotonic information integration throughout the duration of a syntactically coherent or semantically coherent constituent, reflected by monotonic changes in neural activity. In the current model, information integration occurs in a hierarchical manner, with information being compressed at the boundary of each linguistic level. According to this model, neural activity does not necessarily show a monotonic increase within a long constituent but rather decreases at the boundary of each subordinate unit. The current model is more closely related to the syntactic parsing model employed by Nelson and colleagues (Nelson et al., 2017), which assumes that neural activity will track the number of items in the stack of a syntactic parser. The current model, however, has an explicit ground in neural and psychological representations rather than purely linguistic operations.

In the cue-based memory-retrieval model (Lewis et al., 2006), every word is stored in activated long-term memory after it is recognized. When a word needs to be integrated with a previous word, the brain will search for all possible words in activated long-term memory based on the constraints the word must satisfy, e.g., being a noun. The activated long-term memory will decay so that the activation level of a word depends on when the word is heard or retrieved. The current model and the cue-based memory-retrieval model could potentially describe different aspects of language processing. The current model provides a mechanism to efficiently integrate neighboring linguistic units since the buffered linguistic units can be integrated with subsequent linguistic units without additional memory retrieval. The cue-based memory-retrieval model, in contrast, is likely to provide a mechanism to integrate long-distance relationship between words.

The current model is also closely related to the DORA model, which models predicate-argument relationships using symbolic neural networks (Martin and Doumas, 2017). The DORA model is similar to the top-down predicative model in Fig. 2b. In the DORA model reported by Martin and Doumas (2017), the sentences are coded in the model's long-term memory, which supports the retrieval of the whole sentence upon hearing the first word. Therefore, e.g., in Fig. 2, when a word is activated, all its parent nodes are fully activated. The DORA model does allow the parent node and the child node to simultaneously activate. Another important difference between DORA and the current model is that DORA explicitly models the predicate-argument relationship between words while the current model only focuses on the management of items in the processing buffer and does not specify how the words are integrated.

In summary, here we propose a structure-based memory maintenance model to explain neural tracking of speech constituent structures. After specifying a parsing model and other potential parameters, e.g., capacity of the processing buffer and the memory decay constant, the model can generate explicit predictions about the time course of the neural response to a sentence, which can be empirically tested. Therefore, the model provides a general framework to link linguistic theories with neural and psychological representations of linguistic structures.

**Acknowledgement**

I would like to thank Drs. David Poeppel, Lucia Melloni, and Andrea Martin for discussions.